\documentclass[twocolumn,amsmath,amssymb,prl]{revtex4}
\usepackage{color}
\usepackage[pdftex]{graphicx}
\usepackage{dsfont}

\newcommand{\rr}{\mathbf{r}}

\newcommand{\ee}{\mathbf{e}}
\newcommand{\rrho}{\boldsymbol{\rho}}
\newcommand{\kk}{\mathbf{k}}

\newcommand{\pp}{\mathbf{p}}
\newcommand{\Spin}{\boldsymbol{\Sigma}}

\input epsf
\begin{document}
\title {Meron Ground State of Rashba Spin-Orbit Coupled Dipolar Bosons}
\author{Ryan M. Wilson, Brandon M. Anderson and Charles W. Clark}
\affiliation{Joint Quantum Institute, National Institute of Standards and Technology and University of Maryland, Gaithersburg, MD 20899, USA}
\date{\today}
\begin{abstract}
We study the effects of dipolar interactions on a Bose-Einstein condensate with synthetically generated Rashba spin-orbit coupling.  The dipolar interaction we consider includes terms that couple spin and orbital angular momentum in a way perfectly congruent with the single-particle Rashba coupling.  We show that this internal spin-orbit coupling plays a crucial role in the rich ground state phase diagram of the trapped condensate.  In particular, we predict the emergence of a thermodynamically stable ground state with a meron spin configuration.
\end{abstract}
\maketitle

Cold atoms are proving to be indispensable tools for studying quantum many-body physics, addressing the more venerable questions that persist in the field, while introducing new questions of their own~\cite{Dalfovo99,Giorgini08}.  The recent realization of synthetic gauge fields in systems of cold atoms~\cite{Lin11,Chen2012,Wang2012,Cheuk2012} grants promise that they may be used to study, for example, topological insulators and their attendant phenomena.  While such applications are quite exciting, cold atoms are additionally unique in that they provide an opportunity to simulate models beyond those with canonical counterparts, such as spinful bosons in gauge fields~\cite{Stanescu08,Wu2011halfvortex,Sinha11,Xu12PRA,Sedrakyan2012,Hu12}.

A timely example is the atomic Bose-Einstein condensate (BEC) subject to a synthetic gauge field, which provides an ideal platform to study the interplay of spin-orbit coupling (SOC), interactions and superfluidity~\cite{Dalibard11,Galitski13}.  Recent proposals suggest that atom-laser couplings~\cite{Ruseckas05,Juzeliunas10,Campbell11,Xu12} may be used to generate non-Abelian gauge fields for bosons, resulting in, for example, pure Rashba SOC.  These proposals involve the cyclic coupling of more than two atomic hyperfine levels using Raman lasers to produce a dressed $N$-level, or pseudo spin-$(N-1)/2$ ground state with interactions that depend on the choice of the particular undressed levels and laser coupling scheme used.

\begin{figure}[b]
\includegraphics[width=.9\columnwidth]{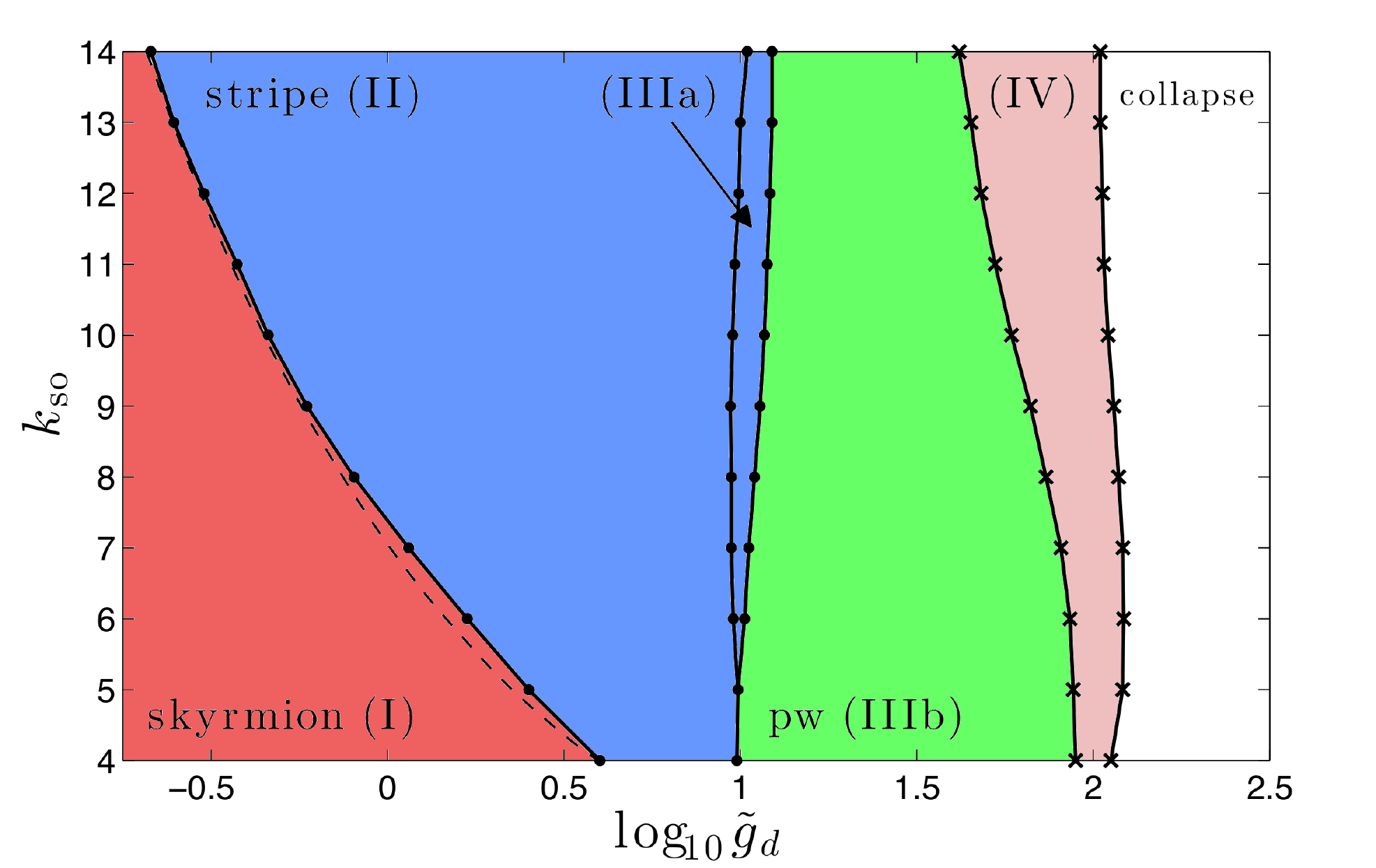} 
\caption{\label{fig:pd} (color online).  Ground state phase diagram of the Rashba spin-orbit coupled Bose-Einstein condensate with pure dipole-dipole interactions (see text for details). }
\end{figure}

In this Letter, we consider a pseudo spin-$\frac{1}{2}$ ($N=2$) system with interactions that are dipolar in nature, as may result from laser-coupling states of highly magnetic atoms such as Dy, Er or Cr, BECs of which have been realized recently~\cite{Griesmaier05,Pasquiou11,Lu11,Aikawa12,Baranov12}.  In contrast to purely contact interactions, the DDI is long range ($\propto 1/r^3$) and anisotropic,  enabling interactions that convert spin with orbital angular momentum in a manner completely parallel with the Rashba SOC.  Employing a mean-field treatment, we explore the effects of such interactions on the harmonically trapped and Rashba spin-orbit coupled BEC and find that the system possesses a rich ground state phase diagram, shown in Fig.~\ref{fig:pd}.  We expect this treatment to give reliable results for a trapped system, where the degeneracy of the single-particle states is lifted relative to the highly degenerate homogeneous system~\cite{Gopal11,Barnett12}.  This phase diagram is in stark contrast to those of BECs without dipolar interactions~\cite{Sinha11,Hu12} and Rashba SOC~\cite{Deng12,Cui13}, and most notably exhibits a discontinuous transition to a meron state.  Interestingly, this meron, which emerges due to the interplay of the single particle Rashba SOC and the internal SOC of the dipolar interaction, exists as a thermodynamically stable ground state.  This is in contrast to other merons that have been studied in the cold atoms context, which emerge under rotation or as quasiparticle excitations of spinor condensates~\cite{Kasamatsu05,Su12,Liu12}.

\emph{Formalism}-- The spin-orbit coupled, interacting Bose gas is described by the many-body Hamiltonian
\begin{equation}
\label{Hfull}
\hat{\mathcal{H}} = \hat{\mathcal{H}}_0 + \hat{\mathcal{H}}_\mathrm{int}.
\end{equation}
The first term in~(\ref{Hfull}) is the single-spinor Hamiltonian including the harmonic trapping potential and the Rashba SOC~\cite{Bychkov84},
\begin{equation}
\label{H0}
\hat{\mathcal{H}}_0 = \int d\rr \, \hat{\Psi}^\dagger(\rr) \left\{ \frac{\pp^2}{2} + \frac{r^2}{2} + k_\mathrm{so} \boldsymbol{\sigma}_\perp \cdot \pp \right\} \hat{\Psi}(\rr),
\end{equation}
where $\hat{\Psi}(\rr) = ( \hat{\psi}_\uparrow(\rr),\hat{\psi}_\downarrow(\rr) )^\mathrm{T}$ is the spinor Bose field annihilation operator for the $\uparrow$ and $\downarrow$ pseudo-spins, $k_\mathrm{so}$ is the SOC strength, $\boldsymbol{\sigma}_\perp = (\sigma_x, \sigma_y, 0)$ and $\sigma_i$ are the Pauli matrices.  In Eq.~(\ref{H0}) and throughout this Letter, we work in dimensionless units by scaling with the appropriate factors of the trap energy $\hbar \omega$ and the trap length $l=\sqrt{\hbar/ M \omega}$, where $M$ is the atomic mass and $\omega$ is the trap frequency.  We define the vector spin-$\frac{1}{2}$ operator $\mathbf{S} = \frac{1}{2}\boldsymbol{\sigma}$.  The Rashba SOC $ \boldsymbol{\sigma}_\perp \cdot \pp$ conserves the total angular momentum projection $ \mathrm{J}_z  =  \mathrm{S}_z  +  \mathrm{L}_z $, with $\mathbf{L} = \rr \times \pp$, but not the spin and orbital projections separately.

The second term in Eq.~(\ref{Hfull}) is the interaction Hamiltonian,
\begin{align}
\label{Hint}
\hat{\mathcal{H}}_\mathrm{int} &= \frac{1}{2} \int d\rr \left\{ \, g \hat{n}^2(\rr) + g_z \hat{\Sigma}_z^2(\rr) - \mu_d \hat{\Spin}(\rr) \cdot \hat{\mathbf{M}}(\rr) \right\}.
\end{align}
The first two terms in~(\ref{Hint}) describe the density-density and spin-spin contact interactions.  Their strengths are given by the couplings $g=4\pi a $ and $g_z= 4\pi a_z $, respectively, where $a$ and $a_z$ are the corresponding $s$-wave scattering lengths.  In~(\ref{Hint}), $\hat{n}(\rr) = \hat{\Psi}^\dagger(\rr) \hat{\Psi}(\rr)$ is the density operator and $ \hat{\Sigma}_z$ is the $z$-component of the spin density operator $\hat{\Spin}(\rr) = \hat{\Psi}^\dagger(\rr) \mathbf{S} \hat{\Psi}(\rr)$.  The third term in~(\ref{Hint}) is the DDI Hamiltonian.  We consider a general pseudo-spin basis where the magnetic dipole operator is simply a magneton times the spin operator, $\mu_d \mathbf{S} $, and write the DDI Hamiltonian as a spin interacting with an effective magnetic field, given by
\begin{equation}
\label{M1}
\hat{\mathbf{M}}(\rr) = \frac{\mu_0 \mu_d}{4\pi} \int d\rr^\prime \frac{3\left( \hat{\Spin}(\rr^\prime) \cdot \mathbf{e}_{\rr-\rr^\prime} \right) \mathbf{e}_{\rr-\rr^\prime} - \hat{\Spin}(\rr^\prime)}{|\rr-\rr^\prime|^3}
\end{equation}
where $\mu_0$ is the magnetic permeability of vacuum and $\mathbf{e}_{\rr-\rr^\prime}$ is the unit vector in the direction of $\rr-\rr^\prime$.  

For dilute bosons at ultracold temperatures, the system is well described by the condensate order parameter $\langle \hat{\Psi}(\rr) \rangle \simeq \sqrt{N_0} \Phi(\rr)$, where $N_0$ is the condensate atom number and $\Phi(\rr)$ is a classical field.  This approximation transforms the Hamiltonian~(\ref{Hfull}) into a classical energy functional, $\langle \hat{\mathcal{H}} \rangle \simeq E[\Phi]$.  Additionally, we utilize the two-dimensional nature of the Rashba SOC in Eq.~(\ref{H0}) by assuming that both spin states occupy the motional ground state in the $z$-direction.  Such an assumption is valid when interaction energies are sufficiently less than the trap energy $\hbar \omega$.  This is justified in retrospect as we find the condensate chemical potential to satisfy $|\mu| < 1$ for all but the very largest values of $\tilde{g}_d$.  We thus posit the separable ansatz $\Phi (\rr) = \phi(\rrho) f(z)$, where $\phi(\rrho) = (\varphi_\uparrow(\rrho),\varphi_\downarrow(\rrho))^\mathrm{T}$ and $f(z) = \exp{[-z^2/2]}/\pi^\frac{1}{4}$, and integrate out the $z$-dependence from $E[\Phi]$ to find an effective energy functional
\begin{align}
\label{Q2DE}
\frac{E[\phi]}{N_0} &= \frac{1}{2} \int d\rrho \left\{ \phi^\dagger(\rrho) \left( \kk^2 + \rho^2 + 2 k_\mathrm{so} \boldsymbol{\sigma}_\perp \cdot \kk \right) \phi(\rrho) \right. \nonumber \\
&+ \left. \tilde{g} n^2(\rrho) + \tilde{g}_z \Sigma_z^2(\rrho)
- \tilde{g}_d \Spin(\rrho) \cdot \mathbf{m}(\rrho) \right\}
\end{align}
where $n(\rrho) = \phi^\dagger(\rrho) \phi(\rrho)$ is normalized to unity and the factors of condensate number are absorbed into the dimensionless coupling constants $\tilde{g} = N_0 \sqrt{8\pi} a $, $\tilde{g}_ z = N_0 \sqrt{8\pi} a_z$ and $\tilde{g}_d = N_0 \mu_0 \mu_d^2 / 3 \sqrt{2\pi} $.  In addition to the global $U(1)$ symmetry, the functional~(\ref{Q2DE}) is $ SO(2)$ symmetric where $SO(2)$ is given by the group $e^{i\mathrm{J}_z \alpha}$ with $\alpha \in [0,2\pi)$.

By the convolution theorem~\cite{Goral02}, the components of the magnetic field generated by the spin distribution can be written as $ \mathrm{m}_i(\rrho) = \mathcal{F}_\mathrm{2D}^{-1} \left[ \tilde{\mathrm{v}}_{ij} (\kk )  \tilde{\Sigma}_j (\kk) \right]$, where $\tilde{\mathrm{v}}_{ij}(\kk) = U_{i \kappa}^{-1} \tilde{\nu}_{\kappa \lambda}(\kk) U_{\lambda j}$ and $\tilde{\nu}_{\kappa \lambda} (\kk)$ is the effective dipole tensor expressed in the spherical basis.  We define the transformation from the cartesian to the spherical basis $\ee_\kappa = U_{\kappa i} \ee_i$, where $\ee_\kappa = (\ee_- , \ee_0 , \ee_+)^\mathrm{T}$, $\ee_{\pm} = \mp (\ee_x \pm i \ee_y)/\sqrt{2}$ and $\ee_0 = \ee_z$.  The non-vanishing elements of this tensor are $\tilde{\nu}_{00} (\kk) = 2 - F\left[ k / \sqrt{2} \right]$, $\tilde{\nu}_{--} (\kk) = \tilde{\nu}^*_{++} (\kk) = \frac{1}{2} F\left[ k / \sqrt{2} \right] e^{2i\alpha_k}$, and $\tilde{\nu}_{-+}(\kk) = \tilde{\nu}_{+-}(\kk) = -\frac{1}{2} \tilde{\nu}_{00}(\kk)$, where $F[x]=3\sqrt{\pi} x e^{x^2} \mathrm{erfc}[x]$, $\mathrm{erfc}$ is the complimentary error function and $\alpha_k$ is the azimuthal angle in $k$-space.

If the spin-operators in the DDI are represented as spin raising and lowering operators, the tensor elements $\tilde{\nu}_{\kappa \lambda}(\kk) $ are seen to provide the momentum dependence for terms $\propto \mathrm{S}_\kappa \mathrm{S}_\lambda$ where $\kappa,\lambda = -,0,+$ and $\pm \leftrightarrow \pm 1$.  The corresponding orbital matrix elements are non-vanishing only when the total $\langle \mathrm{L}_z \rangle$ of the two particles is lowered by $\kappa+\lambda$.  Thus, terms like $\mathrm{S}_\pm \mathrm{S}_\pm$, which we term SOC contributions, raise (lower) the spin angular momentum of two interacting spinful dipoles by $2$ and simultaneously lower (raise) the projection of the orbital angular momentum by $2$, conserving $ \mathrm{J}_z $~\cite{Santos06}.  We refer to the $\langle \mathrm{S}_z \rangle$-  and $\langle \mathrm{L}_z \rangle$-conserving terms $ \propto \mathrm{S}_0 \mathrm{S}_0$ and $\mathrm{S}_\pm \mathrm{S}_\mp$ as direct and spin-exchange contributions, respectively.

\begin{figure*}[t]
\includegraphics[width=.975\textwidth]{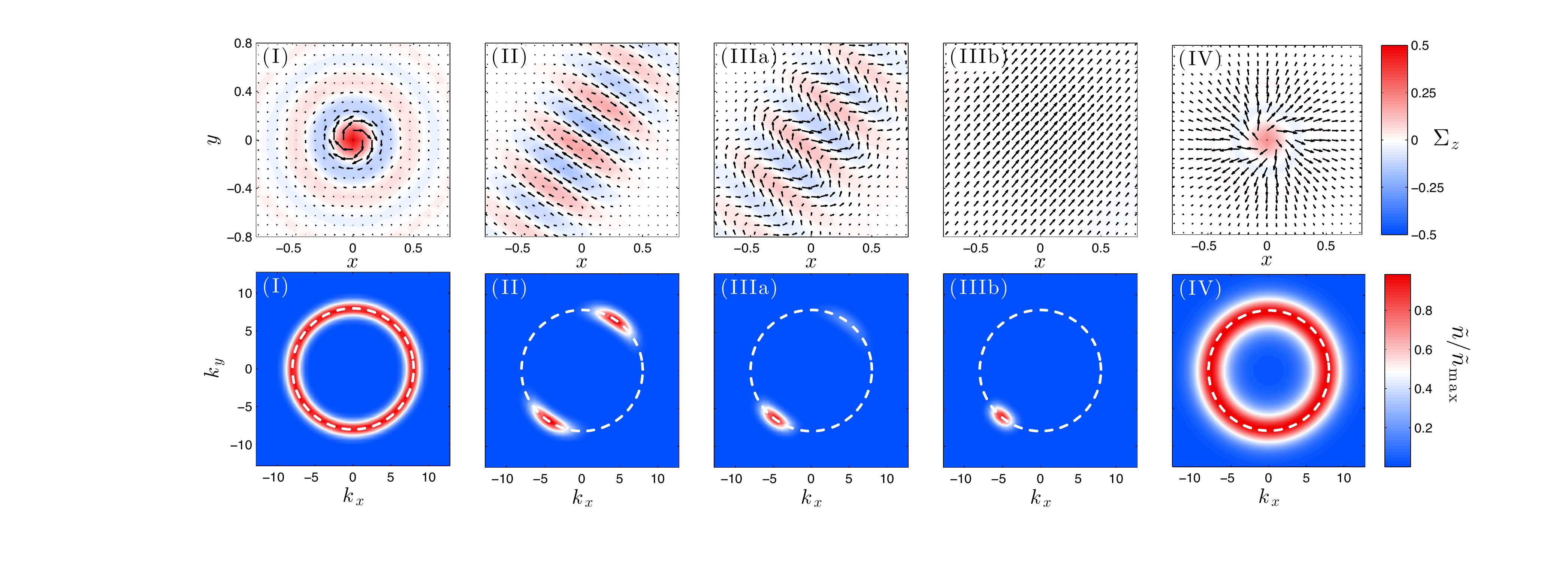}
\vspace{-.3cm}
\caption{\label{fig:phases} (color online).  Examples of spin textures (top row) and $k$-space densities (bottom row) of the ground state phases for $k_\mathrm{so}=8$ and purely dipolar interactions.  They are (I) skyrmion, (II) spin-stripe, (IIIa) finite-pitch spin helix, (IIIb) plane-wave (PW) ferromagnet and (IV) meron.  In this upper row, the shading indicates the $z$-component of the spin density $\Spin(\rrho)$ and the black arrows show the projection of $\Spin(\rrho)$ into the $xy$-plane, and are globally scaled for purposes of visualization.  In the bottom row, the white dashed line indicates the location of the minimum of the Rashba ring.  Images in the bottom row are scaled by $\tilde{n}_\mathrm{max}$, the maximum achieved $k$-space density for each case.}
\end{figure*}

When represented in a spherical basis, the Rashba Hamiltonian is proportional to the operator $e^{i\alpha_k} \mathrm{S}_- + e^{-i\alpha_k} \mathrm{S}_+$.  It is then clear that the action of this Hamiltonian is to raise (lower) the spin angular momentum of a particle and simultaneously lower (raise) the projection of its orbital angular momentum.  This $\mathrm{J}_z$-conserving process is perfectly congruent with the SOC part of the DDI, but at the single-particle level.

\emph{One-body}--  In the absence of the trapping potential, the energy eigenvalues of the in-plane part of the single particle Hamiltonian $\hat{\mathcal{H}}_0$ are given by the dispersions $\varepsilon_\pm = (k\pm k_\mathrm{so}) ^2 / 2$, with the corresponding eigenstates $\phi_{\pm,\kk} (\rrho) = e^{i\kk\cdot\rrho} \chi_{\pm,\kk} $ where $\chi_{\pm,\kk} = (1, \pm e^{i\alpha_k})^\mathrm{T}/\sqrt{2}$.  The lower-energy  branch achieves a minimum along a ring of radius $k = k_\mathrm{so}$, and the ground state manifold is therefore infinitely degenerate with respect to $\alpha_k$.

The presence of a trapping potential breaks this degeneracy, and the single particle states have the $k$-space spinor wave functions $\tilde{ \phi}_{nm} (\kk) = e^{i m \alpha_k} ( \tilde{\varphi}_{\uparrow nm}(k)  , - \tilde{\varphi}_{\downarrow nm}(k) e^{i \alpha_k } )^\mathrm{T}$.   In the limit of large spin orbit coupling $k_\mathrm{so} \gg 1$, the low-lying states with energies much less than $ k_\mathrm{so}^2/2$ have wave functions $\tilde{\varphi}_{\uparrow,nm}(k) \simeq \tilde{\varphi}_{\downarrow,nm}(k)$ that are the 1D harmonic oscillator states centered about $k_\mathrm{so}$.  The corresponding energy eigenvalues are simply those of the radial and angular states in the Rashba ring, $\varepsilon_{nm}  \simeq \frac{1}{2}\left( 2n + 1 + m(m+1) / k_\mathrm{so}^2 \right)$, with the energies $\varepsilon_{nm}$ split by a factor $\sim 1 /k_\mathrm{so}^2$ in a given radial manifold~\cite{Sinha11,Shenoy2011,Hu12,Li2012}.  

\emph{Many-body}--  We find the ground state in the presence of a trapping potential by numerically minimizing the energy functional $E[\phi]$ in Eq.~(\ref{Q2DE})~\cite{note1} for the purely dipolar case $\tilde{g}=\tilde{g}_z=0$.  For small $\tilde{g}_d$, the condensate is well-described by the single-particle ground state ${\phi} \sim {\phi}_{00}$~\cite{Sinha11,Hu12,Ramachandhran12}, shown by the red region (I) in Fig.~\ref{fig:pd}.  The spin texture and the $k$-space density of this state are shown in the first column (I) of Fig.~\ref{fig:phases}.  This state is a half-quantum vortex with $\langle \mathrm{L}_z \rangle / N_0 = \frac{1}{2}$, or a skyrmion with topological charge $Q=1$ where $Q = \int_\Lambda d\rrho \, q(\rrho)$ and $q(\rrho)$ is the topological charge density~\cite{Kasamatsu05},
\begin{equation}
\label{Q}
q(\rrho) = \frac{1}{4\pi} \Spin (\rrho) \cdot \left( \partial_x \Spin(\rrho) \times \partial_y \Spin(\rrho) \right).
\end{equation}
The skyrmion character is revealed by limiting the integral over $q(\rrho)$ to the concentric disk $\Lambda$ with radius $r_\Lambda \simeq j_{0,1} /k_\mathrm{so}$, encompassing one full spin winding, where $j_{0,1}$ is the first zero of the Bessel function $J_0$~\cite{DLMF}. 

In exploring the deviation from the skyrmion ground state with increasing $\tilde{g}_d$, it is insightful to first consider the nature of the ground state in the absence of a radial trapping potential.  We thus consider the class of untrapped states in two dimensions, $\phi = a_1 \phi_{-,\kk} + a_2 \phi_{-,-\kk}$ where $\phi_{-,\kk}$ is a single-particle ground state at a spontaneously chosen value of the angle $\alpha_k$, and $|a_1|^2 + |a_2|^2=1$.  When $a_2=0$, this corresponds to a plane wave (PW) phase, and for $|a_1|=|a_2|$ this corresponds to a spin-stripe phase.  This ansatz thus spans the ground state manifold of the analog system with purely contact interactions~\cite{Wang10}.  We find the condensate to be a PW phase for any finite $k_\mathrm{so}$ and $\tilde{g}_d>0$ in the absence of a radial trapping potential, suggesting that a plane-wave phase will emerge for sufficiently large $\tilde{g}_d$ in the trapped system, though in a form that has no net mass flow~\cite{Ozawa12PRAb}.  A PW state appropriate for a trapped system is a localized Gaussian wave packet on the Rashba ring.  Constructing such a state from the $\phi_{0m}$ wave functions requires coupling in many $m$-states at energy costs $\propto \delta \varepsilon_{0m} = m(m+1)/2k_\mathrm{so}^2$.  The critical value of $\tilde{g}_d$ at which $m>0$ states begin to couple into the ground state can be estimated by finding where the DDI energy of the $\phi_{00}$ state equals $1/k_\mathrm{so}^2$.  This critical $\tilde{g}_d$ is shown by the dashed black line in Fig.~\ref{fig:pd}. 

Instead of transforming from a skyrmion to a PW with increasing $\tilde{g}_d$, the system first exhibits spin-stripe patterning, indicated by the blue region (II) in Fig.~\ref{fig:pd}.  The onset of the stripe phase is seen as the spontaneous breaking of the $SO(2)$ symmetry of the skyrmion in favor of a $C_2 \subset SO(2)$ symmetry  corresponding to $\alpha = 0,\pi$ by coupling in a state $\sim \phi_{0 -2}$.  While a maximally broken symmetry like that of the PW phase is achieved by coupling in $\phi_{01}$, the deviation from the skyrmion spin texture in this case has considerably larger direct and exchange DDI energies than that achieved by coupling in $\phi_{0-2}$.  Thus, the presence of a stripe phase is a finite size effect in this system.

\begin{figure}[t]
\includegraphics[width=.95\columnwidth]{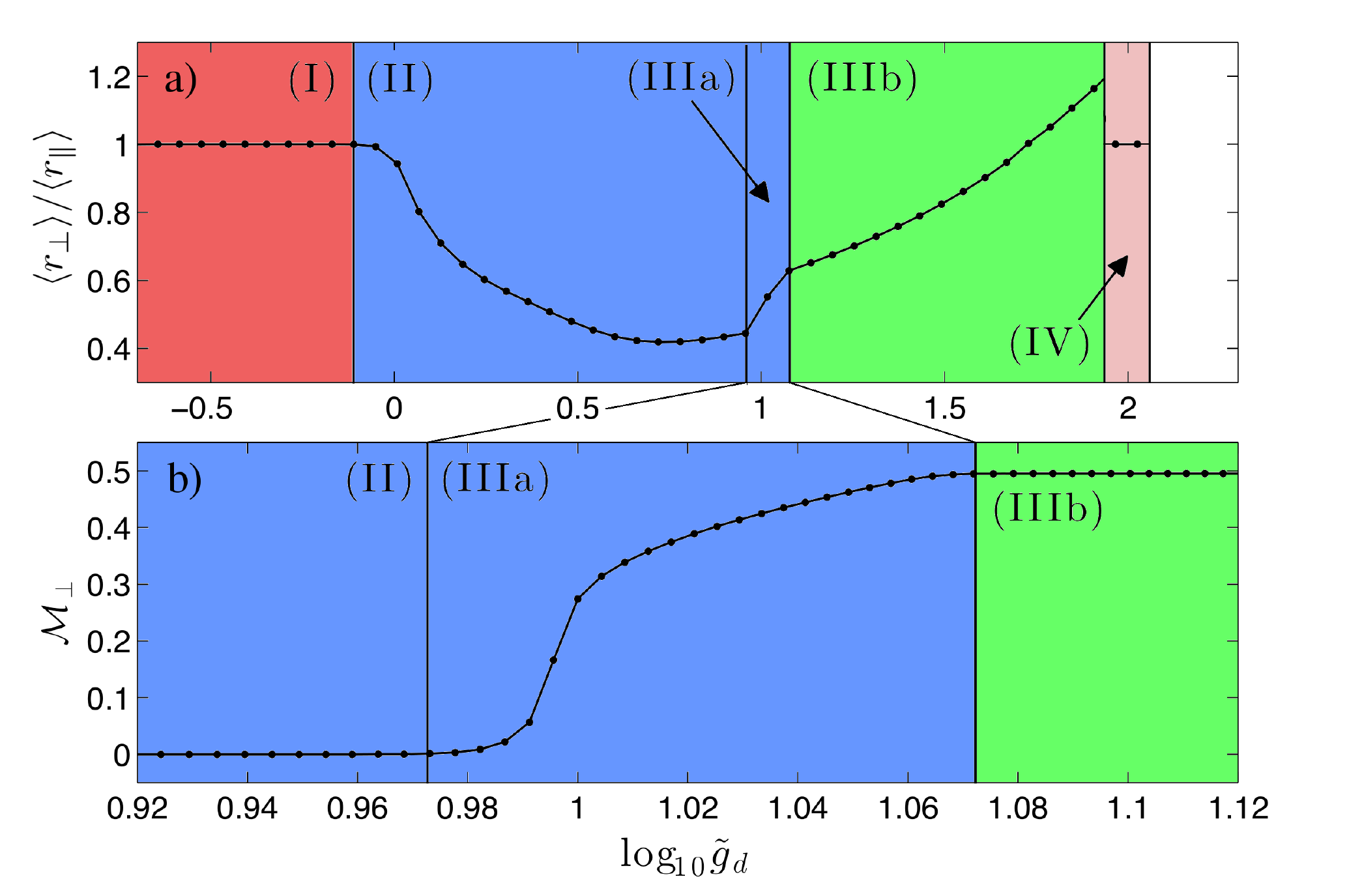} 
\caption{\label{fig:armag} (color online).  a) The real-space condensate aspect ratio for $k_\mathrm{so}=8$.  b)  The in-plane magnetization per particle for $k_\mathrm{so}=10$.  Both cases have $\tilde{g}=\tilde{g}_z=0$ and the Roman numerals refer to phases defined in Fig.~\ref{fig:phases}.   }
\end{figure}

The onset of the stripe phase is continuous with increasing $\tilde{g}_d$ as seen, for example, in the cloud aspect ratio $\langle r_\perp \rangle / \langle r_\parallel \rangle$ where $r_{\perp}$ $(r_\parallel)$ is the direction perpendicular (parallel) to the spin-stripe axis and we define $\langle r_i \rangle \equiv \sqrt{ \langle r_i^2 \rangle} $.  This aspect ratio is plotted in Fig.~\ref{fig:armag}(a) for $k_\mathrm{so}=8$.  In $k$-space, the emergence of stripe ordering is seen as the buildup of two peaks in the $k$-space density at opposite points on the Rashba ring.  The spin texture and $k$-space density of the stripe phase are shown in the second column (II) of Fig.~\ref{fig:phases}.  The striped spin texture has no projection along the spontaneously chosen stripe axis, along which it forms a chiral spin helix.  

Above $\tilde{g}_d \sim 10$, the PW phase emerges as the ground state, shown by region (III) in Fig.~\ref{fig:pd}.  This phase is characterized by a non-vanishing net in-plane  magnetization per particle $\mathcal{M}_\perp = \left( \mathcal{M}_x^2 + \mathcal{M}_y^2 \right)^\frac{1}{2}$ where $\mathcal{M}_i = \int d\rrho \, \Sigma_i(\rrho)$, plotted in Fig.~\ref{fig:armag}(b) for $k_\mathrm{so}=10$.  The condensate is near fully magnetized with $\mathcal{M}_\perp \simeq \frac{1}{2}$ in the green region (IIIb).  The spin texture and $k$-space density of the PW phase with maximal $\mathcal{M}_\perp$ are shown in the fourth column (IIIb) of Fig.~\ref{fig:phases}.  As is clear from the spin texture, the PW phase is in-plane ferromagnetic, and is similar to phases identified in other dipolar Bose fluids~\cite{Yi06,Kawaguchi06,Huhtamaki10}.  This ferromagnetic spin texture characterizes the ground state in the limit of vanishing $k_\mathrm{so}$.  

The stripe-PW transition is continuous and supports an intermediate phase (IIIa), shown by the thin blue region in Fig.~\ref{fig:pd}.  The spin texture and $k$-space density of this phase are shown in the third column (IIIa) of Fig.~\ref{fig:phases}.  The $k$-space density resembles that of the spin-stripe, but with unequal peak amplitudes.  The resulting spin texture is therefore a chiral spin helix, but with a finite spin projection onto the stripe axis, or a finite pitch.  This finite-pitch spin helix has a finite but non-maximal in-plane magnetization, as seen in Fig.~\ref{fig:armag}(b).

For the $\tilde{g}_d$ considered thus far, the ground state phases, to an excellent approximation, can be decomposed into the single-particle states of the lowest radial band $\phi_{0m}$.  A condensate composed of two such states $\phi = a_m \phi_{0m} + a_{m^\prime} \phi_{0 m^\prime}$ with the relative phase $\gamma= \arg a_m a_{m^\prime} $ gives a contribution to the DDI energy that scales as $\cos^4 \frac{\gamma}{2}$.  Coupling in states from higher radial bands introduces DDI energy scalings like $\cos \gamma$ and $\cos 2\gamma$, allowing for the DDI energy to change sign.  Indeed, for sufficiently large $\tilde{g}_d$, the ground state couples in states with radial quantum number $n>0$ to achieve a negative contribution from the SOC part of the DDI energy.  This state is a meron with an $SO(2)$ symmetry and fractional topological charge $Q=\frac{1}{2}$ for $r_\Lambda = \infty$~\cite{Mermin76}. 

Merons are objects of fundamental importance in high energy field theories~\cite{Actor79} and in solid state systems, where they play a central role in the quantization of spin-Hall conductance~\cite{Bernevig06}, in the Kosterlitz-Thouless transition of binary 2D electron gases~\cite{Moon95} and emerge in ground state configurations of quantum Hall ferromagnets~\cite{Brey96,Milo09}.  In the cold atoms context, they have been predicted in rotating two-component BECs~\cite{Kasamatsu05} and in quenched spin-1 BECs~\cite{Su12,Liu12}, though not as a stationary, thermodynamically stable ground state, as we predict here.  This distinction is significant, making this meron more robust to experimental manipulation and measurement, and more akin to those relevant in solid state systems.  The meron exists in the pink region (IV) in Fig.~\ref{fig:pd}, and its spin texture and $k$-space density are shown in the fifth column (IV) of Fig.~\ref{fig:phases}.   The PW-meron transition is discontinuous, reflecting the topologically distinct character of these phases.  Beyond a large critical $\tilde{g}_d$, no stable mean-field solution exists, and the condensate collapses under an excessively attractive SOC part of the DDI.

\emph{Discussion}--  As mentioned above, current proposals for generating spin-orbit coupled, pseudo spin-$\frac{1}{2}$ atomic ground states are generally strongly dependent on the given atomic species and exact Raman laser coupling scheme used~\cite{Ruseckas05,Juzeliunas10,Campbell11,Xu12,Zhang12}.  However, coupling schemes that are symmetric across an atomic spin manifold, like that proposed in~\cite{Deng12}, will result in a DDI like that in Eq.~(\ref{Hint}).  

The phases we predict here emerge at experimentally relevant parameters.  For example, consider $^{164}$Dy~\cite{Lu11} with a magneton $\mu_d = 10 \mu_\mathrm{B}$ where $\mu_\mathrm{B}$ is the Bohr magneton.  A trap with $\omega = 2\pi \times 200 \, \mathrm{Hz}$ and a particle number $N_0 = 20\times 10^3$ gives a DDI coupling $\tilde{g}_d \simeq 100$, which is deep in the parameter space of the meron phase, and corresponds to a very dilute condensate.  For the SOC strengths we consider here, $4 \lesssim k_\mathrm{so} \lesssim 14$, the BEC critical temperature is predicted to be about two-thirds of that without Rashba SOC ($T_c \sim 1 \, \mu\mathrm{K}$)~\cite{Hu12b}, putting this system well within the reach of modern experiments.  While our mean-field description neglects quantum and thermal fluctuations, at sufficiently low temperatures we expect the fluctuations to be only weakly perturbative in the dilute regime~\cite{OzawaPRL12}.  Regarding experimental detection, the momentum-space densities and spin textures of the phases presented here are readily observable through spin-resolved time-of-flight measurements~\cite{Lin11}.

\emph{Conclusion}--  In this Letter, we consider the effects of the dipole-dipole interaction on Rashba spin-orbit coupled BECs.  The competition between the single-particle SOC and the DDI leads to a rich and unique phase diagram involving novel spin textures, phase transitions and emergent topological states.  Of particular interest is the contribution from the internal SOC of the DDI, which is perfectly parallel with the Rashba SOC and thus plays an important role in the ground state physics and phase diagram, leading to the presence of a meron ground state.

\emph{Note added}-- The results presented in this Letter are complemented by the recent Ref.~\cite{Gopal13arXiv}, where pure density-density dipolar interactions are considered for a Rashba spin-orbit coupled condensate in a homogeneous quasi-two-dimensional geometry.

\emph{Acknowledgements}-- R. M. W. acknowledges support from an NRC postdoctoral fellowship.  This work was partially supported by the NSF under the Physics Frontiers Center Grant PHY-0822671, and the ARO  Atomtronics MURI.


\begin{thebibliography}{53}
\bibitem[{not()}]{note1}
\bibinfo{note}{We obtain reasonably accurate results when using the asymptotic
  momentum-independent values of $\tilde{\nu}_{\alpha \beta}(\kk)$ for larger
  $k_\mathrm{so}$.}
\end{thebibliography}
\end{document}